\documentclass[times,twocolumn,final]{elsarticle}

\usepackage{framed,multirow}

\usepackage{amssymb}
\usepackage{amsmath}
\usepackage{latexsym}

\usepackage{url}
\usepackage{xcolor}

\usepackage{hyperref}
\usepackage{cleveref}
\usepackage{algorithm}  
\usepackage{algpseudocode}

\usepackage{booktabs}
\definecolor{newcolor}{rgb}{.8,.349,.1}
\usepackage{geometry}
\journal{Computerized Medical Imaging and Graphics}

\begin{document}


\begin{frontmatter}

\title{Complementary consistency semi-supervised learning for 3D left atrial image segmentation}%

\author[1]{Hejun Huang}

\author[1,2]{Zuguo Chen\corref{cor1}}
\cortext[cor1]{Corresponding author.}
\ead{zg.chen@hnust.edu.cn}
\author[1]{Chaoyang Chen}

\author[1]{Ming Lu}

\author[1]{Ying Zou}
\address[1]{School of Information and Electrical Engineering, Hunan University of Science and Technology, Xiangtan 411201, China}
\address[2]{Shenzhen Institute of Advanced Technology, Chinese Academy of Sciences, Shenzhen 518055, China}


\begin{abstract}
A network based on complementary consistency training, called CC-Net, has been proposed for semi-supervised left atrium image segmentation. CC-Net efficiently utilizes unlabeled data from the perspective of complementary information to address the problem of limited ability of existing semi-supervised segmentation algorithms to extract information from unlabeled data. The complementary symmetric structure of CC-Net includes a main model and two auxiliary models. The complementary model inter-perturbations between the main and auxiliary models force consistency to form complementary consistency. The complementary information obtained by the two auxiliary models helps the main model to effectively focus on ambiguous areas, while enforcing consistency between the models is advantageous in obtaining decision boundaries with low uncertainty. CC-Net has been validated on two public datasets. In the case of specific proportions of labeled data, compared with current advanced algorithms, CC-Net has the best semi-supervised segmentation performance.
Our code is publicly available at \url{https://github.com/Cuthbert-Huang/CC-Net}.

\end{abstract}

\begin{keyword}
\texttt Complementary consistency\sep Semi-supervised segmentation\sep Complementary auxiliary models\sep Uncertainty
\end{keyword}

\end{frontmatter}


\section{Introduction}
\label{sec1}
Atrial fibrillation (AF) is the most common arrhythmia and has a significant impact on global mortality, becoming one of the major burdens of global healthcare \citep{Guglielmo_Mutimodality2019}. The structure of the left atrium (LA) is necessary information for clinicians to diagnose and treat atrial fibrillation \citep{Ikenouchi_The2021}. Traditional methods of manually segmenting LA have great limitations due to their strong empirical dependence and susceptibility to errors \citep{Xiong_A_global2021}. Deep learning-based methods have been developed for automatic segmentation of LA. For example, a multi-task learning framework was constructed to share features among tasks and achieve accurate segmentation \citep{Chen_Muti-task2019}. An attention-based hierarchical aggregation network (HAANet) was proposed, using hierarchical aggregation to enhance the network's feature fusion ability and attention mechanisms to improve the extraction of effective features \citep{Li_Attention2019}. These supervised learning methods have demonstrated good segmentation performance but require a large amount of annotated data for training. 3D medical image annotation data is scarce due to difficulty and high cost of annotation. Therefore, how to achieve good segmentation performance with fewer annotated data remains a pressing problem.

Semi-supervised learning typically refers to a method that utilizes a large amount of unlabeled data and a small amount of labeled data to jointly learn, and is better suited for scenarios where labeled data is difficult to obtain \citep{van_a_survey2020}. Semi-supervised learning is particularly effective in the segmentation of the left atrium in 3D. Based on the mean teacher model, \citet{Yu_Uncertainty-aware2019} uses an uncertainty map to guide the student model to gradually learn reliable information from the teacher model, resulting in good left atrium segmentation results. \citet{Li_Shape2020} uses a signed distance map regression to introduce shape and position prior information, while using a discriminator as a regularization term to enhance segmentation stability. \citet{Luo_Semi2021} recognizes the disturbance between regression and prediction tasks, and constructs a bi-task consistency loss through task conversion to learn unlabeled data, enhancing the model's generalization ability. Although the above works have achieved good left atrium segmentation performance, they have not been able to learn information from difficult areas of unlabeled data. MC-Net+ \citep{Wu_Mutual_consistency2022} constructs the mutual consistency between three different upsampling decoders, generating low-entropy predictions for areas of uncertainty, and achieves effective results. However, the probability maps generated by different upsampling methods only contain conservative learnable information, and due to the shared encoder, the learnable information is weakened as the training progresses, leading to MC-Net+ being unable to obtain correct segmentation results in critical areas of uncertainty (see the comparison between our method and MC-Net+ in the Performance on the LA dataset section \ref{subsec4-2} for details).

\begin{figure*}[!t]
	\centering
	\includegraphics[scale=.8]{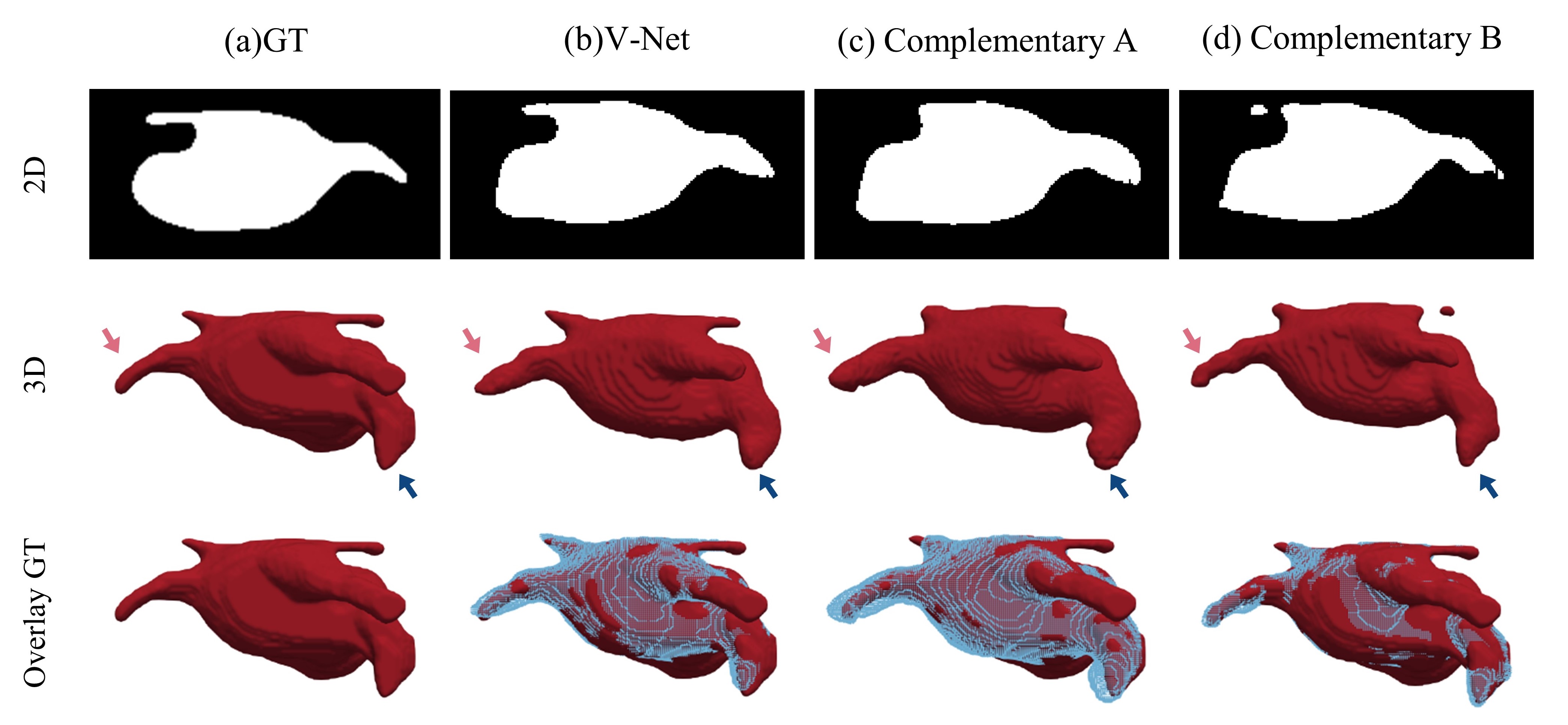}
	\caption{Comparison of segmentation results of Complementary A, Complementary B, and V-Net after training with 10\% labeled data, where Complementary A is obtained from the second and fourth layers of the V-Net decoder without skip-connection, and Complementary B is obtained from the first and third layers of the V-Net decoder without skip-connection.}
	\label{fig1}
\end{figure*}

The article argues that accurate segmentation is achieved by combining high-level semantic information with high-resolution detail information. Focusing more on high-level semantic information means expanding the boundaries of the deterministic segmentation region, that is, reducing false negative rates. Focusing more on high-resolution detail information means reducing the uncertainty of correct segmentation boundaries, that is, increasing true positive rates. Can the model be adjusted to focus more on high-level semantic information or more on high-resolution detail information to obtain probability maps with rich learnable information? Skip connections play an important role in V-Net, helping to restore high-resolution detail information lost during encoding during upsampling \citep{V-net2016}. Complementary A and Complementary B are obtained by changing whether the skip connection is used in a certain layer of the V-Net decoder. Figure \ref{fig1} compares the segmentation results on the LA dataset after training with 10\% labeled data using Complementary A, Complementary B, and V-Net. Complementary A gives up some high-resolution detail information and focuses more on high-level semantic information. The second row of Figure\ref{fig1}(c) indicates that Complementary A has wider segmentation boundaries in the challenging branch area (indicated by the arrow in the figure). The third row of Figure \ref{fig1}(c) clearly shows that the segmentation region of Complementary A basically wraps around the true label. Complementary B gives up some high-level semantic information and focuses more on high-resolution detail information. The second row of Figure \ref{fig1}(d) indicates that Complementary B has more reliable segmentation boundaries in the challenging branch area (indicated by the arrow in the figure). The third row of Figure \ref{fig1}(d) clearly shows that the segmentation region of Complementary B is basically wrapped by the true label. The results above show that there is rich learnable information between the probability maps generated by the Complementary A and Complementary B models.

Therefore, this paper proposes a new network based on complementary consistency training by using V-Net as the main model and constructing two complementary auxiliary models. The two auxiliary models form a complementary symmetric structure by changing whether to use skip connections in a certain layer of the V-Net decoder. Drawing on cross-pseudo-supervision \citep{Chen_Semi2021} and mutual consistency \citep{Wu_Semi2021}, a sharpening function is used to convert the probability maps generated by the two auxiliary models into pseudo-labels to strengthen the training of the main model. At the same time, the high-quality probability maps generated by the main model are also converted into pseudo-labels to guide the training of the auxiliary models. The perturbation between the main model and the auxiliary models forms a complementary consistency training. The Dice loss is used as the supervised loss for labeled input data, and the model consistency regularization loss is used as the unsupervised loss for all input data. After training, only the main model is used for testing, greatly reducing the number of network parameters during testing while achieving fine segmentation results.Consequently, the contributions and novelty of this paper are summarised as follows:
\begin{itemize}
	\item This paper utilizes two complementary auxiliary models that are constructed by alternating the use of skip connections. This creates a model disturbance from a complementary information perspective, effectively utilizing unlabeled data.
	\item The use of model consistency methods allows the main model to learn complementary information from the auxiliary models.
	\item An independent encoder structure is proposed for complementary consistency learning.
\end{itemize}
This method is validated on two public datasets and the results show that it effectively increases the utilization of unlabeled data and achieves excellent semi-supervised segmentation performance.
\section{Related work}
\label{sec2}

\subsection{Semi-supervised segmentation}
\label{subsec2-1}
Due to the high cost and difficulty in obtaining labeled data in segmentation tasks, semi-supervised segmentation has been vigorously developed to learn useful information from unlabeled data as much as possible. \citet{Zhai_ASS-GAN2022} defined two asymmetric generators and a discriminator for adversarial learning, in which the discriminator filters the mask generated by the generators to provide supervised information for the network to learn from unlabeled data. \citet{Xiao_Efficient2022} added a teacher model that combines CNN and Transformer structures on the basis of the mean teacher model, aiming to guide the student model to learn more information. \citet{Zhang_Discriminative2022} proposed a dual correction method to improve the quality of pseudo labels and obtain better segmentation performance. This paper constructs a complementary auxiliary model to help the main model explore the ambiguous area of unlabeled data, and the complementary consistency between the main model and the auxiliary model effectively learns from unlabeled data.

\subsection{Consistency regularization}
\label{subsec2-2}
Consistency regularization is a common and effective method in semi-supervised learning. To address inherent perturbations among related tasks, \citet{Luo_Semi2021} introduced the dual-task consistency between the segmentation map derived from the level set and directly predicted segmentation map. \citet{Liu_A2022} added a classification model on top of the segmentation model and constructed a contrastive consistency loss using the class vectors obtained from the classification model. \citet{Wang_DC-net2022} added spatial context perturbations to the input on top of model-level perturbations, resulting in a dual-consistency segmentation network. \citet{Hu_Semi2022} embedded self-attention in the mean teacher model and encouraged the attention maps to remain consistent across feature layers, forming attention-guided consistency. In the cross-modal domain, \citet{Chen_MASS2022} used cross-modal consistency between non-paired CT and MRI to learn modality-independent information. \citet{Ouali_Semi2020} proposed cross-consistency to enforce consistency between the primary decoder and auxiliary decoder in decision-making for low-density regions. To address perturbations between different levels of image enhancement, \citet{Zhong_Pixel2021} introduced pixel-wise contrastive consistency based on label consistency property and contrastive feature space property between pixels. Inspired by mutual consistency, our approach uses complementary consistency based on model consistency to enable the main model to learn complementary information from the auxiliary model.

\subsection{Multi-view training}
\label{subsec2-3}
The purpose of multi-view training is to utilize the redundant information between different views to improve learning efficiency \citep{Yan_muti-view_review2021}. \citet{Dong_Tri-net2018} constructs three divergent models and generates pseudo-labels using a voting mechanism. With a large amount of unlabeled data and the introduction of noise, this method yields good results. \citet{Xia_Uncertainty-aware2020} performs multi-view collaborative training by rotating the input and using the uncertainty estimates from each view to obtain accurate segmentation. This method is simple to implement, but requires a relatively large number of views to achieve ideal performance, leading to redundant learning. \citet{Zheng_Uncertainty2022} splits labeled data into complementary subsets and trains two models with each subset, effectively improving the network's ability to explore ambiguous areas. Our approach constructs two complementary auxiliary models to guide the main model's attention to ambiguous areas from two complementary views. Additionally, leveraging the multi-view information leads to low-entropy predictions.

\subsection{Uncertainty estimation}
\label{subsec2-4}
Uncertainty estimation helps reduce the randomness of predictions, and is crucial for learning reliable information from unlabeled data \citep{Chen_Uncertainty2022}. In \citep{Yu_Uncertainty-aware2019}, Monte Carlo sampling is used to obtain uncertainty maps from the teacher model, which guides the student model to gradually acquire reliable information. \citet{Zheng_Uncertainty2022} uses uncertainty maps as weights for the loss to learn high-confidence information. \citet{Wu_Mutual_consistency2022} constructs three different upsampling decoders and uses mutual learning to obtain low-uncertainty predictions. \citet{Wang_tripled-uncertainty} uses a triple uncertainty-guided framework, allowing the student model to obtain more reliable knowledge from the teacher model for all three tasks. Our method reduces uncertainty in complementary information through mutual learning between the complementary auxiliary models and the main model.

\section{Materials and Methods}
\label{sec3}
\subsection{Dataset and pre-processing}
\label{subsec3-1}
This article evaluates the proposed method using the LA dataset and the Pancreas-CT dataset.

The LA dataset \citep{Xiong_A_global2021} used in this study is obtained from the 2018 Atrial Segmentation Challenge and consists of 154 3D LGE-MRIs from 60 diagnosed atrial fibrillation patients. Each 3D LGE-MRI scan has an isotropic resolution of $0.625\times 0.625\times 0.625m{{m}^{3}}$ and spatial dimensions of $576\times 576\times 88$ or $640\times 640\times 88$ pixels. The segmentation labels were manually obtained by three trained observers and stored in NRRD format. Since only 100 labeled images were available, this study followed the settings of \citep{Yu_Uncertainty-aware2019,Li_Shape2020,Luo_Semi2021,Wu_Mutual_consistency2022}, where the 100 images were split into 80 for training and 20 for validation. The proposed method's performance was compared with other methods using the same validation set.

The Pancreas-CT dataset \citep{clark2013cancer} consists of 82 3D contrast-enhanced CT scans from 53 male and 27 female patients. In 2020, cases \#25 and \#70 were found to be duplicates of case \#2 with minor cropping differences and were removed from the dataset. The CT scans have a resolution of $512\times 512$ pixels, with pixel sizes and slice thicknesses between 1.5-2.5 millimeters. Following [31], we resampled the voxel sizes to a uniform isotropic resolution of $1.0\times 1.0\times 1.0m{{m}^{3}}$ and truncated the Hounsfield Units (HU) to the range of [-125, 275]. We used 60 samples for training and reported the performance on the remaining 20 samples.
\begin{figure*}[!t]
	\centering
	\includegraphics[scale=.7]{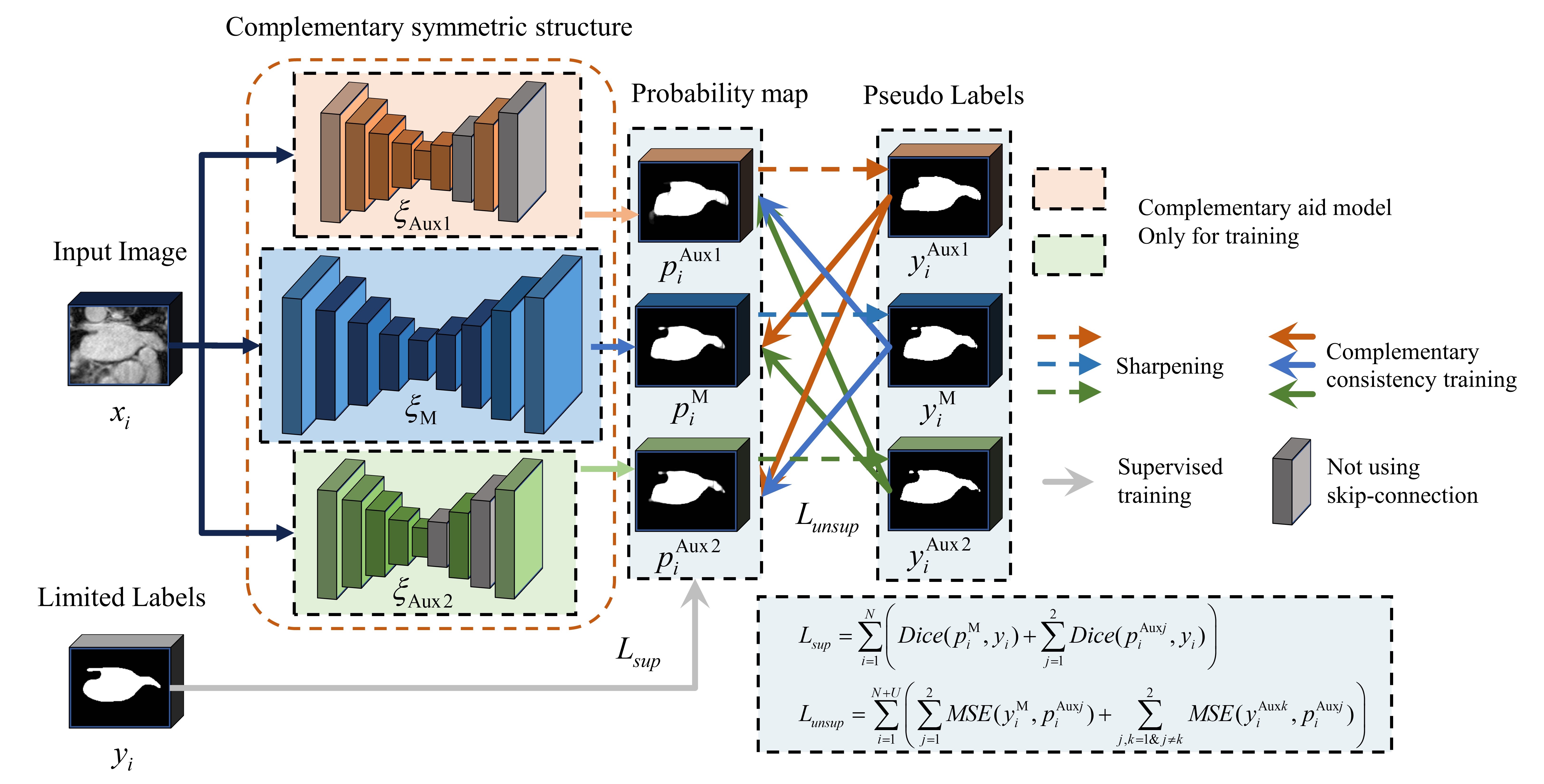}
	\caption{Overall architecture of CC-Net. The complementary symmetric structure consists of a main model ${{\xi }_{M}}$ and two auxiliary models ${{\xi }_{\operatorname{Aux}1}}$, ${{\xi }_{\operatorname{Aux}2}}$. The two auxiliary models alternate using skip connections to form complementary model-level perturbations. The input image ${{x}_{i}}$ is simultaneously fed into the three models, resulting in corresponding probability maps $p_{i}^{\text{M}}$, $p_{i}^{\text{Aux1}}$ and $p_{i}^{\text{Aux2}}$. The sharpened probability maps $y_{i}^{\text{M}}$, $y_{i}^{\text{Aux1}}$ and $y_{i}^{\text{Aux2}}$ are obtained by applying a sharpening function to $p_{i}^{\text{M}}$, $p_{i}^{\text{Aux1}}$ and $p_{i}^{\text{Aux2}}$. The limited number of real labels y guides the model's output and generates the supervised loss ${{L}_{sup}}$, while the pseudo label metrics model output generates the unsupervised loss ${{L}_{unsup}}$.}
	\label{fig2}
\end{figure*}

\subsection{Network overall architecture}
\label{subsec3-2}
The overall architecture of our proposed method is illustrated in Figure \ref{fig2}. It consists of a main model and two auxiliary models. The main model employs V-Net \citep{V-net2016}, and the encoders of the two auxiliary models are identical to that of the main model. The decoders of the auxiliary models are constructed as a complementary symmetric structure using skip connections in an interleaved manner. Specifically, the second and fourth layers of the decoder in auxiliary model 1 do not have skip connections, while the first and third layers of the decoder in auxiliary model 2 do not have skip connections. The segmentation network takes 3D medical images as input. If the input is labeled data, the main and auxiliary models are supervisedly trained using real labels. If the input is unlabeled data, the two auxiliary models learn complementary information of the same input through the complementary symmetric structure to effectively utilize the unlabeled data. The probability maps generated by the main model and auxiliary models are sharpened to obtain pseudo labels. The complementary information learned by one of the auxiliary models is used to guide the training of the main model and the other auxiliary model through the pseudo label. At the same time, the pseudo labels generated by the main model guide the training of both auxiliary models. This leads to the formation of a complementary consistency training network (CC-Net) consisting of a main model and two complementary auxiliary models. It is noteworthy that the two auxiliary models are only used for training, and the main model is used for inference, which greatly improves the efficiency of inference. The next two subsections will describe in detail the complementary symmetric structure and complementary consistency training.

\begin{figure}[!t]
	\centering
	\includegraphics[scale=.4]{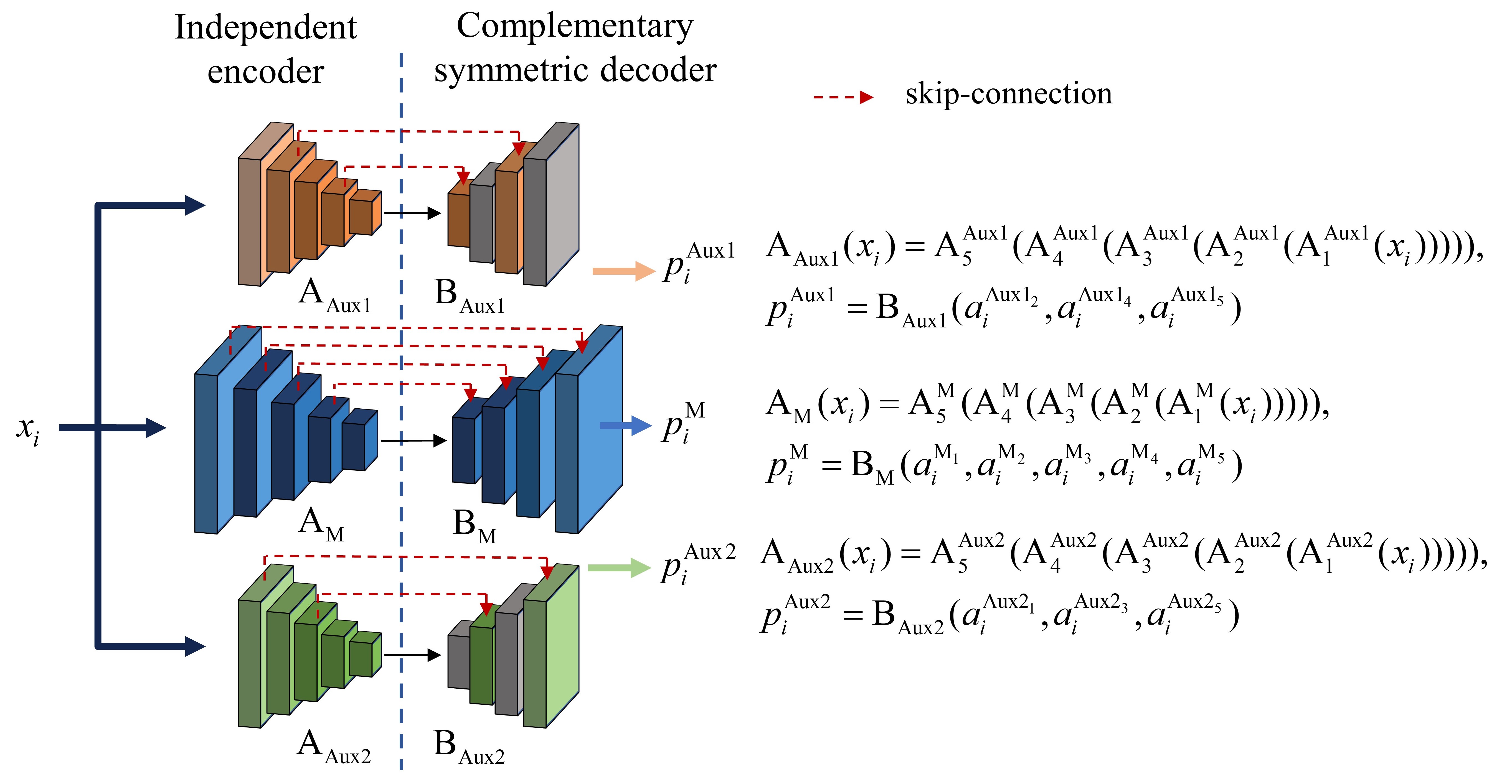}
	\caption{Overview of the complementary symmetric structure. The encoder structures of the main model and two auxiliary models are identical but independent. The four decoding blocks of the main model all use skip connections. For auxiliary model 1, skip connections are used on the first and third decoding blocks. For auxiliary model 2, skip connections are used on the second and fourth decoding blocks.}
	\label{fig3}
\end{figure}
\subsection{Complementary symmetric structure}
\label{subsec3-3}
Inspired by \citep{Wang_tripled-uncertainty}, this paper proposes a method that utilizes two complementary auxiliary models to learn complementary information and effectively utilize unannotated data. As shown in Figure \ref{fig3}, the main model and auxiliary models adopt independent encoders. The three encoders have the same structure, denoted as ${{\text{A}}_{\text{M}}}$, ${{\text{A}}_{\text{Aux1}}}$ and ${{\text{A}}_{\text{Aux2}}}$ respectively. For the input x, each encoder has 5 encoding blocks and can produce 5 outputs after 5 encodings.

\begin{equation}
	\begin{aligned}
	& a_{i}^{\operatorname{Aux}{{1}_{j}}}=\operatorname{A}_{j}^{\text{Aux1}}(\operatorname{A}_{j-1}^{\text{Aux1}}({{x}_{i}}))), \\ 
	& a_{i}^{{{\operatorname{M}}_{j}}}=\operatorname{A}_{j}^{\operatorname{M}}(\operatorname{A}_{j-1}^{\operatorname{M}}({{x}_{i}}))), \\ 
	& a_{i}^{\operatorname{Aux}{{2}_{j}}}=\operatorname{A}_{j}^{\text{Aux2}}(\operatorname{A}_{j-1}^{\text{Aux2}}({{x}_{i}}))) \\ 
	\label{eq1}
	\end{aligned}
\end{equation}
where $a_{i}^{\text{Aux}{{1}_{j}}}$ represents the $j\text{-th}(0<j\le 5)$ result produced by the $i\text{-th}$ data input to the first auxiliary model encoder, $a_{i}^{{{\text{M}}_{j}}}$ represents the $j\text{-th}(0<j\le 5)$ result produced by the $i\text{-th}$ data input to the main model encoder, and $a_{i}^{\text{Aux}{{\text{2}}_{j}}}$ represents the $j\text{-th}(0<j\le 5)$ result produced by the $i\text{-th}$ data input to the second auxiliary model encoder. $\operatorname{A}_{j}^{\text{Aux}1}$ represents the $j\text{-th}(0<j\le 5)$ encoding block of the first auxiliary model, $\operatorname{A}_{j}^{\operatorname{M}}$ represents the $j\text{-th}(0<j\le 5)$ encoding block of the main model encoder, and $\operatorname{A}_{j}^{\text{Aux2}}$ represents the $j\text{-th}(0<j\le 5)$ encoding block of the second auxiliary model encoder.

The decoders of the auxiliary models form complementary model-level perturbations. ${{\text{B}}_{\text{M}}}$, ${{\text{B}}_{\text{Aux1}}}$, and ${{\text{B}}_{\text{Aux2}}}$ representing the decoders of the main model and the two auxiliary models, respectively. Decoder layers 2 and 4 of auxiliary model 1 do not use skip-connection, and the inputs to decoder ${{\text{B}}_{\text{Aux1}}}$ are the final output $a_{i}^{\text{Aux}{{1}_{5}}}$ and intermediate outputs $a_{i}^{\text{Aux}{{1}_{2}}}$ and $a_{i}^{\text{Aux}{{1}_{4}}}$ of encoder ${{\text{A}}_{\text{Aux1}}}$. Thus the probability map of the auxiliary model 1 is as follows.
\begin{equation}
	\begin{aligned}
		p_{i}^{\text{Aux1}}={{\text{B} }_{\text{Aux1}}}(a_{i}^{\text{Aux}{{1}_{2}}},a_{i}^{\text{Aux}{{1}_{4}}},a_{i}^{\text{Aux}{{1}_{5}}})
	\label{eq2}
	\end{aligned}
\end{equation}
The first auxiliary model sacrifices some high-resolution feature fusion in favor of focusing more on high-level semantic information. By expanding the segmentation boundaries to include more voxels that should be segmented, it reduces the number of false negatives. Decoder layers 1 and 3 of auxiliary model 2 do not use skip-connection, and the inputs to decoder ${{\text{B}}_{\text{Aux2}}}$ are the final output $a_{i}^{\text{Aux}{{\text{2}}_{5}}}$ and intermediate outputs $a_{i}^{\text{Aux}{{\text{2}}_{1}}}$ and $a_{i}^{\text{Aux}{{\text{2}}_{3}}}$ of encoder ${{\text{A} }_{\text{Aux2}}}$. Thus the probability map of the auxiliary model 2 is as follows.
\begin{equation}
	\begin{aligned}
		p_{i}^{\text{Aux2}}={{\text{B} }_{\text{Aux2}}}(a_{i}^{\text{Aux}{{\text{2}}_{1}}},a_{i}^{\text{Aux}{{\text{2}}_{3}}},a_{i}^{\text{Aux}{{\text{2}}_{5}}})
	\label{eq3}
	\end{aligned}
\end{equation}
Auxiliary model 2 sacrifices some of the low-resolution feature fusion to focus more on high-resolution pixel information. By optimizing the segmentation boundaries, it reduces the number of incorrectly segmented voxels and increases the true positive rate. The main model, on the other hand, uses the complete V-Net, with all four decoding blocks using skip connections, thus forming a simple model-level perturbation with the auxiliary models. The probability map of the main model is as follows.
\begin{equation}
	\begin{aligned}
		p_{i}^{\text{M}}={{\text{B} }_{\text{M}}}(a_{i}^{{{\text{M}}_{1}}},a_{i}^{{{\text{M}}_{2}}},a_{i}^{{{\text{M}}_{3}}},a_{i}^{{{\text{M}}_{4}}},a_{i}^{{{\text{M}}_{5}}})
	\label{eq4}
	\end{aligned}
\end{equation}

Auxiliary model 1 focuses more on high-level semantic information, while Auxiliary model 2 focuses more on high-resolution pixel information. By simple configuration, the two auxiliary models are inclined in different directions while retaining compatible information, forming a complementary symmetrical structure. The complementary symmetrical setting in this paper uses a simple complementary symmetrical skip connection. The two auxiliary models not only learn effective complementary information, but also form minimal model-level perturbations with the main model, ensuring that the main model effectively accepts guidance.

\subsection{Complementary consistency training}
\label{subsec3-4}
If the training set used in this study has $N$ labeled data and $U$ unlabeled data, and if $N\ll U$, then the labeled data set is denoted as ${{D}_{L}}=\left\{ {{x}_{i}},{{y}_{i}} \right\}_{i=1}^{N}$ and the unlabeled data set is ${{D}_{U}}=\left\{ {{x}_{i}} \right\}_{i=N+1}^{N+U}$, where ${{x}_{i}}\in {{\mathbb{R}}^{H\times W\times D}}$ is the input image and ${{y}_{i}}\in {{\mathbb{R}}^{H\times W\times D}}$ is the corresponding label. We construct two auxiliary models ${{\xi }_{\text{Aux1}}}$ and ${{\xi }_{\text{Aux2}}}$ for the main model ${{\xi }_{\text{M}}}$. The input image x is input to both the main and auxiliary models, and the respective probability maps $p_{i}^{\text{M}}$, $p_{i}^{\text{Aux1}}$ and $p_{i}^{\text{Aux2}}$ are obtained from \cref{eq1,eq2,eq3,eq4}.

Based on the clustering assumption, the decision boundary is located in the low-density area. To reduce the influence of pixels that are easily misclassified, a sharpening function is used to transform the probability map $p$ into a pseudo-label $y*$, which can strengthen the entropy minimization constraint. The sharpening function \citep{Wu_Semi2021} is defined as follows:
\begin{equation}
	\begin{aligned}
		y*=\frac{{{p}^{1/T}}}{{{p}^{1/T}}+{{(1-p)}^{1/T}}}
		\label{eq5}
	\end{aligned}
\end{equation}

where, $T$ is a hyperparameter that controls the temperature of the sharpening function. An appropriate $T$ can effectively enhance the entropy minimization constraint and avoid introducing excessive noise and interference. The probability maps of the main model and auxiliary models are sharpened to obtain pseudo labels $y_{i}^{\text{M}}$, $y_{i}^{\text{Aux1}}$, and $y_{i}^{\text{Aux2}}$, respectively.

For the labeled input image ${{x}_{i}}\in {{D}_{L}}$, we use the label ${{y}_{i}}\in {{D}_{L}}$to guide the learning of the main and auxiliary models. The supervised loss uses the common dice loss. Supervised losses are defined as:
\begin{equation}
	\begin{aligned}
		{{L}_{sup}}=\sum\limits_{i=1}^{N}{\left( Dice(p_{i}^{\text{M}},{{y}_{i}})+\sum\limits_{j=1}^{2}{Dice(p_{i}^{\text{Aux}j},{{y}_{i}})} \right)}
		\label{eq6}
	\end{aligned}
\end{equation}
where, $p_{i}^{\text{M}}$ denotes the probability map generated by the primary model for the $i\text{-th}$ input, $p_{i}^{\text{Aux}j}$ denotes the probability map generated by the $j\text{-th}$ auxiliary model for the $i\text{-th}$ input, and $Dice$ represents the dice loss. For all input images, the complementary auxiliary model generates probability maps with rich learnable information, and the pseudo-labels of the complementary auxiliary model guide the training of the main model. Due to the lack of overall segmentation capability of a single auxiliary model, training needs to be guided by the pseudo label generated by the main model and another auxiliary model. The unsupervised loss uses the mean square error (MSE) between the pseudo label and the probability map. Unsupervised loss is defined as:
\begin{equation}
	\begin{aligned}
		{{L}_{unsup}}=\sum\limits_{i=1}^{N+U}{\left( \sum\limits_{j=1}^{2}{MSE(y_{i}^{\text{M}},p_{i}^{\text{Aux}j})}+\sum\limits_{j,k=1\And j\ne k}^{2}{MSE(y_{i}^{\text{Aux}k},p_{i}^{\text{Aux}j})} \right)}
	\label{eq7}
	\end{aligned}
\end{equation}
where $y_{i}^{\text{M}}$ denotes the pseudo label generated by the main model for the $i\text{-th}$ input, $p_{i}^{\text{Aux}j}$ denotes the probability map generated by the $j\text{-th}$ auxiliary model for the $i\text{-th}$ input, $y_{i}^{\text{Aux}k}$ denotes the pseudo label generated by the $k\text{-th}$ auxiliary model for the $i\text{-th}$ input, and $MSE$ denotes the mean square error. The final loss is the sum of the supervised loss and the unsupervised loss. Therefore, the final loss is defined as:
\begin{equation}
	\begin{aligned}
		{{L}_{total}}={{\lambda }_{s}}{{L}_{sup}}+{{\lambda }_{u}}{{L}_{unsup}}
	\label{eq8}
	\end{aligned}
\end{equation}
where ${{\lambda }_{s}}$ is the weight coefficient of supervised loss, ${{\lambda }_{u}}$ is the weight coefficient of unsupervised loss, ${{L}_{sup}}$ is used for labeling input images only, and ${{L}_{un\sup }}$ is used for all input images. According to \citep{Yu_Uncertainty-aware2019,Luo_Semi2021}, ${{\lambda }_{u}}$ use a time-independent Gaussian warming-up function. ${{\lambda }_{s}}$ is set to 0.3 to enhance the effect of complementary consistency training. The semi-supervised learning algorithm based on complementary consistency training is shown in Algorithm \ref{alg1}.
\begin{algorithm}[h]  
	\caption{Complementary consistency-based semi-supervised learning,}  
	\label{alg1}  
	\begin{algorithmic}[1]
		\Require  
		${{x}_{i}}\in {{D}_{L}}+{{D}_{U}}$,${{y}_{i}}\in {{D}_{L}}$
		\Ensure 
		main model parameters $\theta $ 
		\State Initialize iteration = 0, max\_iteration = 10000
		\State ${{\xi }_{\text{M}}}(x)$ = main model with parameters $\theta $
		\State ${{\xi }_{\text{Aux1}}}(x)$ = auxiliary model 1 with parameters ${{\theta }_{1}}$
		\State ${{\xi }_{\text{Aux2}}}(x)$ = auxiliary model 2 with parameters ${{\theta }_{2}}$
		\While {iteration $<$ max\_iteration}
		\State Calculate the probability maps $p_{i}^{\text{M}}$, $p_{i}^{\text{Aux1}}$, and $p_{i}^{\text{Aux2}}$ of the primary and secondary models according to \cref{eq1,eq2,eq3,eq4}.
		\If {${{x}_{i}}\in {{D}_{L}}$} 
		\State Calculate the supervised loss ${{L}_{sup}}$ according to equation \eqref{eq6}.
		\EndIf
		\State Calculate the respective pseudo labels $y_{i}^{\text{M}}$, $y_{i}^{\text{Aux1}}$, and $y_{i}^{\text{Aux2}}$ of the primary and secondary models according to equation \eqref{eq5}.
		\State Calculate the unsupervised loss ${{L}_{unsup}}$ according to equation \eqref{eq7}.
		\State Calculate the final loss ${{L}_{total}}$ according to equation \eqref{eq8}.
		\State Calculate the gradient of the loss function ${{L}_{total}}$ and update the main model parameters $\theta $ and auxiliary model parameters ${{\theta }_{1}}$ and ${{\theta }_{2}}$ , by backpropagation.
		\State iteration = iteration + 1
		\EndWhile
		\State \textbf{return} $\theta $
	\end{algorithmic}  
\end{algorithm}

\begin{table*}[htbp]
	\centering
	\caption{Comparison of quantitative results of various methods on the LA dataset (where * indicates data from MC-Net+ \citep{Wu_Mutual_consistency2022}). The model parameter count (Para.) and multiple accumulation operation counts (MACs) are measured during model inference.}
	\begin{tabular}{llllllll}
		\toprule
		\multirow{2}[4]{*}{Method} & \multicolumn{2}{l}{\#Scans used} & \multicolumn{3}{l}{Metrics} & \multicolumn{2}{l}{Complexity} \\
		\cmidrule{2-8}          & Labeled & Unlabeled & Dice(\%)↑ & 95HD(voxel)↓ & ASD(voxel)↓ & Para.(M) & MACs(G) \\
		\midrule
		V-Net* & 8(10\%) & 0     & 78.57 & 21.2  & 6.07  & 9.44  & 41.45 \\
		V-Net* & 16(20\%) & 0     & 86.96 & 11.85 & 3.22  & 9.44  & 41.45 \\
		V-Net* & 80(All) & 0     & 91.62 & 5.4   & 1.64  & 9.44  & 41.45 \\
		CC-Net(ours) & 8(10\%) & 0     & 69.55 & 22.62 & 5.99  & 9.44  & 41.45 \\
		CC-Net(ours) & 16(20\%) & 0     & 76.93 & 18.19 & 2.55  & 9.44  & 41.45 \\
		CC-Net(ours) & 80(All) & 0     & 91.21 & 5.82  & 1.43  & 9.44  & 41.45 \\
		\midrule
		\midrule
		UA-MT \citep{Yu_Uncertainty-aware2019} (MICCAI) & 8(10\%) & 72(90\%) & 85.69 & 16.14 & 4.39  & 9.44  & 41.45 \\
		SASSNet \citep{Li_Shape2020} (MICCAI) & 8(10\%) & 72(90\%) & 86.8  & 14.57 & 4.11  & 9.44  & 41.45 \\
		DTC \citep{Luo_Semi2021} (AAAI) & 8(10\%) & 72(90\%) & 87.43 & 8.37  & 2.4   & 9.44  & 41.45 \\
		MC-Net \citep{Wu_Semi2021} (MICCAI) & 8(10\%) & 72(90\%) & 87.5  & 11.28 & 2.3   & 12.35 & 83.88 \\
		MC-Net+ \citep{Wu_Mutual_consistency2022} (MIA) & 8(10\%) & 72(90\%) & 88.9  & 8.02  & 1.91  & 9.44  & 41.45 \\
		\midrule
		\textbf{CC-Net(ours)} & 8(10\%) & 72(90\%) & \textbf{89.82} & \textbf{7.03} & \textbf{1.81} & 9.44  & 41.45 \\
		\midrule
		\midrule
		UA-MT \citep{Yu_Uncertainty-aware2019} (MICCAI) & 16(20\%) & 64(80\%) & 88.87 & 7.32  & 2.26  & 9.44  & 41.45 \\
		SASSNet \citep{Li_Shape2020} (MICCAI) & 16(20\%) & 64(80\%) & 89.17 & 8.57  & 2.86  & 9.44  & 41.45 \\
		DTC \citep{Luo_Semi2021} (AAAI) & 16(20\%) & 64(80\%) & 89.43 & 7.39  & 2.12  & 9.44  & 41.45 \\
		MC-Net \citep{Wu_Semi2021} (MICCAI) & 16(20\%) & 64(80\%) & 90.12 & 8.07  & 1.99  & 12.35 & 83.88 \\
		MC-Net+ \citep{Wu_Mutual_consistency2022} (MIA) & 16(20\%) & 64(80\%) & 91.05 & 5.81  & 1.69  & 9.44  & 41.45 \\
		\midrule
		\textbf{CC-Net(ours)} & 16(20\%) & 64(80\%) & \textbf{91.27} & \textbf{5.75} & \textbf{1.54} & 9.44  & 41.45 \\
		\bottomrule
	\end{tabular}%
	\label{tab1}%
\end{table*}%

\begin{figure*}[htbp]
	\centering
	\includegraphics[scale=.8]{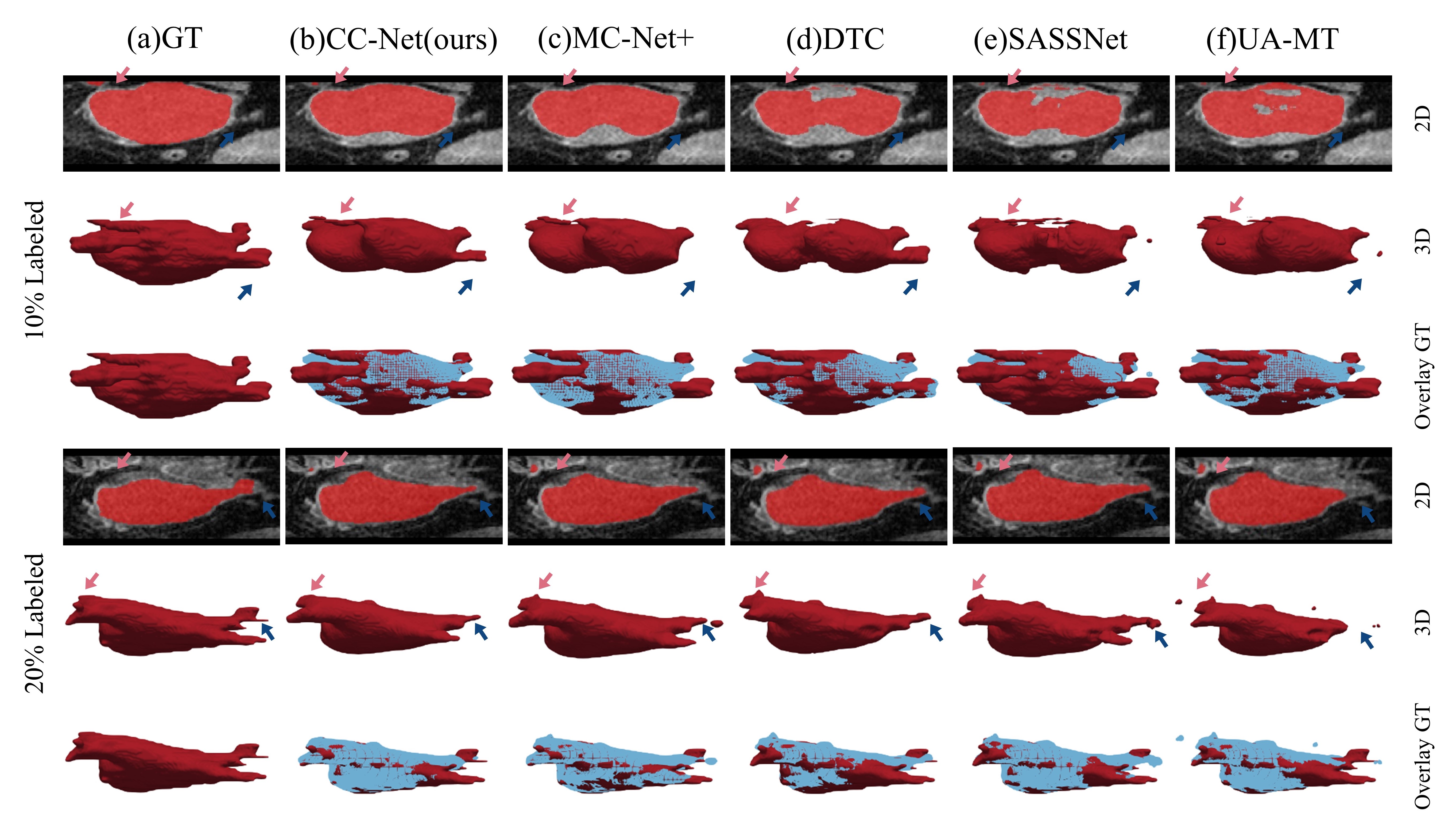}
	\caption{Visual comparison of the segmentation results on LA dataset using semi-supervised training. (a) Ground Truth, (b) CC-Net (proposed method), (c) MC-Net+ \citep{Wu_Mutual_consistency2022}, (d) DTC \citep{Luo_Semi2021}, (e) SASSNet \citep{Li_Shape2020}, (f) UA-MT \citep{Yu_Uncertainty-aware2019}. The first and second rows show 2D and 3D views, respectively. The third row shows the 3D wireframe (in blue) of the corresponding algorithm overlaid with the surface of GT.}
	\label{fig4}
\end{figure*}

\begin{table*}[htbp]
	\centering
	\caption{Comparison of quantitative results of various methods on the Pancreas-CT dataset (where * indicates data from MC-Net+ \citep{Wu_Mutual_consistency2022}). The model parameter count (Para.) and multiple accumulation operation counts (MACs) are measured during model inference.}
	\begin{tabular}{llllllll}
		\toprule
		\multirow{2}[4]{*}{Method} & \multicolumn{2}{l}{\#Scans used} & \multicolumn{3}{l}{Metrics} & \multicolumn{2}{l}{Complexity} \\
		\cmidrule{2-8}          & Labeled & Unlabeled & Dice(\%)↑ & 95HD(voxel)↓ & ASD(voxel)↓ & Para.(M) & MACs(G) \\
		\midrule
		V-Net* & 6(10\%) & 0     & 67.58 & 21.2  & 6.07  & 9.44  & 41.45 \\
		V-Net* & 12(20\%) & 0     & 76.33 & 8.12  & 2.09  & 9.44  & 41.45 \\
		V-Net* & 60(All) & 0     & 83.19 & 5.45  & 1.23  & 9.44  & 41.45 \\
		CC-Net(ours) & 6(10\%) & 0     & 67.21 & 14.19 & 2.56  & 9.44  & 41.45 \\
		CC-Net(ours) & 12(20\%) & 0     & 74.46 & 10.66 & 1.41  & 9.44  & 41.45 \\
		CC-Net(ours) & 60(All) & 0     & 83.21 & 5.11  & 1.18  & 9.44  & 41.45 \\
		\midrule
		\midrule
		UA-MT \citep{Yu_Uncertainty-aware2019} (MICCAI) & 6(10\%) & 54(90\%) & 68.09 & \textbf{12.93} & 3.92  & 9.44  & 41.45 \\
		DTC \citep{Luo_Semi2021} (AAAI) & 6(10\%) & 54(90\%) & 61.06 & 29.17 & \textbf{3.07} & 9.44  & 41.45 \\
		MC-Net+ \citep{Wu_Mutual_consistency2022} (MIA) & 6(10\%) & 54(90\%) & 68.3  & 14.4  & 4.81  & 9.44  & 41.45 \\
		\midrule
		\textbf{CC-Net(ours)} & 6(10\%) & 54(90\%) & \textbf{68.32} & 15.45 & 3.21  & 9.44  & 41.45 \\
		\midrule
		\midrule
		UA-MT \citep{Yu_Uncertainty-aware2019} (MICCAI) & 12(20\%) & 48(80\%) & 72.17 & 11.64 & 3.36  & 9.44  & 41.45 \\
		DTC \citep{Luo_Semi2021} (AAAI) & 12(20\%) & 48(80\%) & 72.2  & 13.81 & 1.84  & 9.44  & 41.45 \\
		MC-Net+ \citep{Wu_Mutual_consistency2022} (MIA) & 12(20\%) & 48(80\%) & 73.32 & 8.28  & 1.18  & 9.44  & 41.45 \\
		\midrule
		\textbf{CC-Net(ours)} & 12(20\%) & 48(80\%) & \textbf{77.8} & \textbf{8.28} & \textbf{1.17} & 9.44  & 41.45 \\
		\midrule
		\midrule
		UA-MT \citep{Yu_Uncertainty-aware2019} (MICCAI) & 18(30\%) & 42(70\%) & 74.56 & \textbf{9.3} & \textbf{1.82} & 9.44  & 41.45 \\
		DTC \citep{Luo_Semi2021} (AAAI) & 18(30\%) & 42(70\%) & 72.73 & 13.04 & 1.87  & 9.44  & 41.45 \\
		MC-Net+ \citep{Wu_Mutual_consistency2022} (MIA) & 18(30\%) & 42(70\%) & 77.73 & 17.56 & 3.8   & 9.44  & 41.45 \\
		\midrule
		\textbf{CC-Net(ours)} & 18(30\%) & 42(70\%) & \textbf{78.69} & 15.35 & 4.16  & 9.44  & 41.45 \\
		\midrule
		\midrule
		UA-MT \citep{Yu_Uncertainty-aware2019} (MICCAI) & 30(50\%) & 30(50\%) & 79.9  & 7.94  & 1.89  & 9.44  & 41.45 \\
		DTC \citep{Luo_Semi2021} (AAAI) & 30(50\%) & 30(50\%) & 76.25 & 10.39 & 1.33  & 9.44  & 41.45 \\
		MC-Net+ \citep{Wu_Mutual_consistency2022} (MIA) & 30(50\%) & 30(50\%) & 82.04 & 6.02  & 1.28  & 9.44  & 41.45 \\
		\midrule
		\textbf{CC-Net(ours)} & 30(50\%) & 30(50\%) & \textbf{82.57} & \textbf{5.73} & \textbf{1.27} & 9.44  & 41.45 \\
		\bottomrule
	\end{tabular}%
	\label{tab2}%
\end{table*}%

\begin{figure*}[htbp]
	\centering
	\includegraphics[scale=.8]{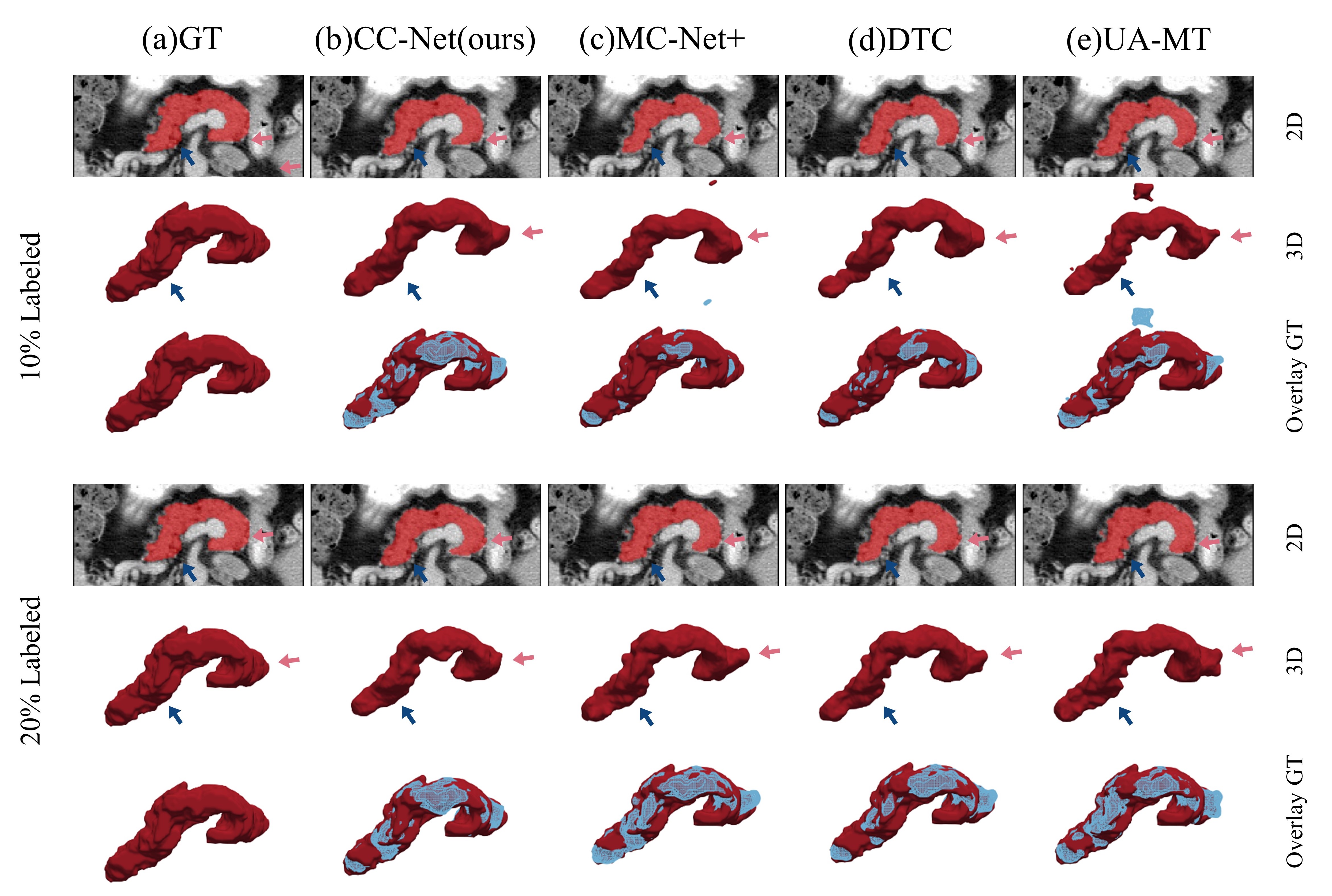}
	\caption{Visual comparison of the segmentation results on Pancreas-CT dataset using semi-supervised training. (a) Ground Truth, (b) CC-Net (proposed method), (c) MC-Net+ \citep{Wu_Mutual_consistency2022}, (d) DTC \citep{Luo_Semi2021}, (e) UA-MT \citep{Yu_Uncertainty-aware2019}. The first and second rows show 2D and 3D views, respectively. The third row shows the 3D wireframe (in blue) of the corresponding algorithm overlaid with the surface of GT.}
	\label{fig5}
\end{figure*}

\section{Experiments and results}
\label{sec4}
\subsection{Implementing details and evaluation indicators}
\label{subsec4-1}
The experiments in this paper were conducted on a single NVIDIA Tesla V100 using Pytorch 1.9.1+cuda11.1 and Python 3.6.5. The SGD optimizer was used with a learning rate of 0.01 and weight decay for 10K iterations. The pseudo-labels were generated using a sharpening function with a temperature constant of T=0.1. All experiments were conducted with a random seed of 1337. During training, random cropping, rotation, and flipping were used for data augmentation. The LA dataset was processed into patches of size $112\times 112\times 80$, and the Pancreas-CT dataset was processed into patches of size $96\times 96\times 96$. During testing, the LA dataset was cropped into patches of size $112\times 112\times 80$ with a step size of 18 in the x and y directions and 4 in the z direction. The Pancreas-CT dataset was cropped into patches of size $96\times 96\times 96$ with a step size of 16 in all directions. The final results were obtained by assembling the predicted results of each block. The two auxiliary models were not used during testing, and the results were based solely on the output of the main model. No post-processing was used in any of the experiments during testing.

According to \citet{Yu_Uncertainty-aware2019}, the Dice score \citep{eelbode2020optimization} was used to evaluate the internal segmentation, and the average surface distance (ASD) and 95th percentile Hausdorff distance (95HD) were used to evaluate the edge segmentation. Due to the different isotropic resolutions of the two datasets, voxels were used as the distance unit.

\subsection{Performance on the LA dataset}
\label{subsec4-2}
Table \ref{tab1} provided in the paper shows a quantitative comparison of the results obtained on the LA dataset using semi-supervised learning compared to fully supervised learning using V-Net and CC-Net. The proposed method achieved a Dice score of 89.82\%, a 95HD of 7.03 voxels, and an ASD of 1.81 voxels with only 10\% of the data being labeled. When compared to CC-Net's fully supervised training, the Dice score increased from 69.55\% to 89.82\%, an improvement of 20.27\%. When compared to V-Net's fully supervised training, the Dice score increased from 78.57\% to 89.82\%, an improvement of 11.25\%. CC-Net trained with 20\% labeled data performed similarly to V-Net trained with all labeled data, with a Dice score of 91.27\% compared to 91.62\%, a 95HD of 5.75 voxels compared to 5.40 voxels, and an ASD of 1.54 voxels compared to 1.64 voxels. The results demonstrate that the proposed semi-supervised learning method can effectively utilize unlabeled data and achieve results comparable to fully supervised learning when labeled data is limited.

The paper also provides visual segmentation results in Figure \ref{fig4}, showing that the proposed method can effectively segment challenging regions such as small branches and local protrusions. The paper compares the segmentation results of different algorithms and notes that DTC has a strong ability to segment branches, but the interior segmentation is rough. MC-Net+ has a good overall segmentation effect, but the branches are often ignored. SASSNet has many errors in the internal segmentation, while UA-MT is not sensitive to the segmentation of branches. The proposed method's segmentation results are closer to the ground truth when compared to other advanced algorithms, as shown in the blue wireframe and 3D label fusion images.

\subsection{Performance on the Pancreas-CT dataset}
\label{subsec4-3}
To further validate the effectiveness of our proposed method, we extended the segmentation task to the Pancreas-CT dataset. Table \ref{tab2} shows the comparison results of our method and three advanced algorithms. Pancreas segmentation is a challenging task, and CC-Net achieves similar segmentation performance to V-Net when trained with 50\% labeled data in semi-supervised learning. Moreover, in all five semi-supervised learning settings, CC-Net achieves the best segmentation performance. Figure \ref{fig5} provides a visual comparison of the results on the Pancreas-CT dataset, and from the perspectives of 2D, 3D, and overlay GT, especially in the areas indicated by arrows, the segmentation results of our proposed method are closer to the true labels compared to other advanced algorithms.

\subsection{Analysis of the stability of test results}
\label{subsec4-4}
\begin{figure*}[htbp]
	\centering
	\includegraphics[scale=.8]{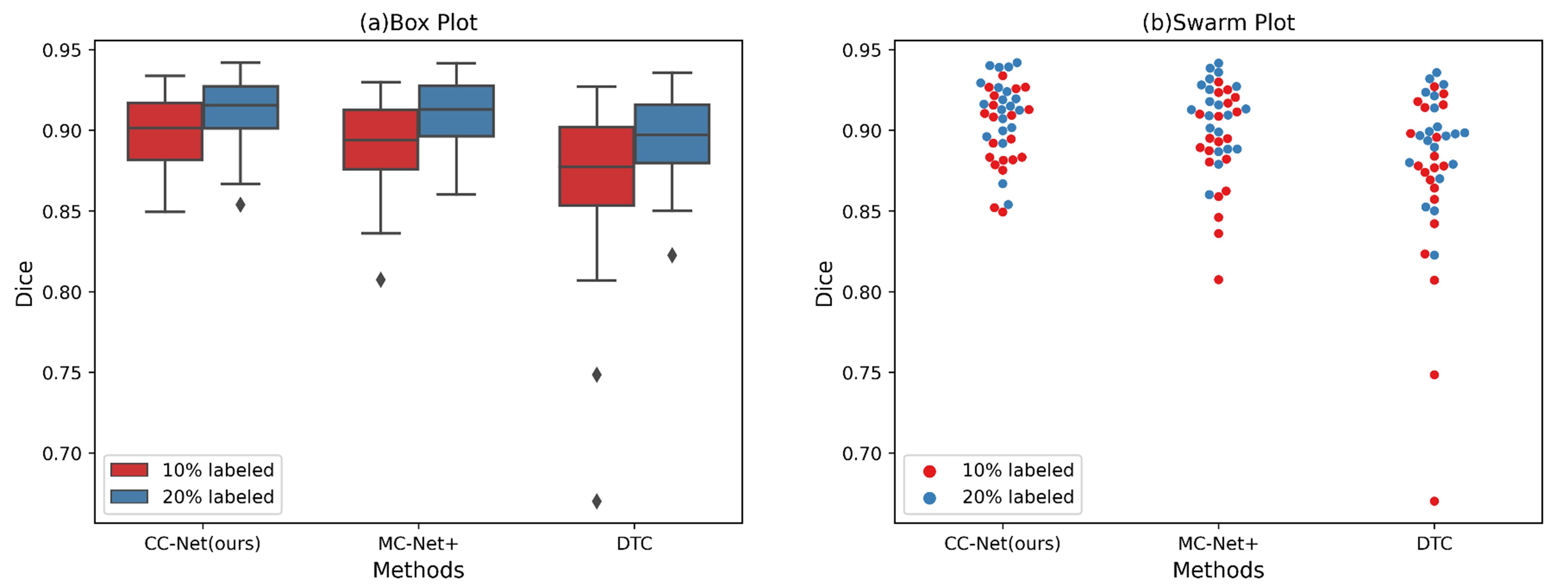}
	\caption{The distribution results of 20 test samples on the LA dataset are compared among CC-Net, MC-Net+ \citep{Wu_Mutual_consistency2022}, and DTC \citep{Luo_Semi2021}. The test results are shown using (a) box plot and (b) swarm plot.}
	\label{fig6}
\end{figure*}

Table \ref{tab3} provides a descriptive statistics of the test results for 20 LA dataset samples. For the 10\% labeled data setting, CC-Net outperforms the other two methods in terms of mean, median, and standard deviation. For the 20\% labeled data setting, CC-Net and MC-Net+ have similar standard deviations, but CC-Net has higher mean and median values compared to MC-Net+. Figure 5 shows the boxplots and scatterplots of Dice scores for our proposed method and other state-of-the-art methods on the 20 test samples of the LA dataset. The red color indicates the results trained with 10\% labeled data. The boxplot of Dice scores for CC-Net (Figure \ref{fig6}(a)) has a smaller interquartile range (IQR) than MC-Net+ and DTC, and the median is higher than MC-Net+ and DTC. Both MC-Net+ and DTC have outliers in their boxplots. Combined with the scatterplot (Figure \ref{fig6}(b)), it is evident that CC-Net has a more concentrated distribution of test results, and can achieve better performance on challenging samples. The blue color indicates the results trained with 20\% labeled data. The boxplot of Dice scores for CC-Net (Figure \ref{fig6}(a)) has a smaller IQR than MC-Net+ and DTC, and the median is higher than MC-Net+ and DTC, despite having one outlier. Combined with the scatterplot (Figure \ref{fig6}(b)), it is easy to see that CC-Net overall outperforms MC-Net+ and DTC in terms of test results. In summary, for these 20 test samples, our proposed method has a more concentrated distribution of test results and better overall performance compared to the other two methods, demonstrating a stronger stability of the test results.

\begin{table*}[htbp]
	\centering
	\caption{Descriptive statistics of the LA dataset test results under two semi-supervised training settings. N represents the sample size, Max represents the maximum value, Min represents the minimum value, SD represents the standard deviation, Med represents the median, and CV represents the coefficient of variation. Bold indicates the best result.}
	\begin{tabular}{lllllllll}
		\toprule
		Methods & Labeled & N     & Max   & Min   & Mean  & SD    & Med   & CV \\
		\midrule
		CC-Net(ours) & 10\%  & 20    & \textbf{0.934} & \textbf{0.849} & \textbf{0.898} & \textbf{0.025} & \textbf{0.901} & \textbf{0.0274684} \\
		MC-Net+ \citep{Wu_Mutual_consistency2022} & 10\%  & 20    & 0.93  & 0.807 & 0.889 & 0.033 & 0.894 & 0.036853717 \\
		DTC \citep{Luo_Semi2021}   & 10\%  & 20    & 0.927 & 0.67  & 0.863 & 0.063 & 0.877 & 0.072833469 \\
		\midrule
		CC-Net(ours) & 20\%  & 20    & \textbf{0.942} & 0.854 & \textbf{0.913} & 0.023 & \textbf{0.916} & 0.025491147 \\
		MC-Net+ \citep{Wu_Mutual_consistency2022} & 20\%  & 20    & 0.942 & \textbf{0.86} & 0.911 & \textbf{0.022} & 0.913 & \textbf{0.023975588} \\
		DTC \citep{Luo_Semi2021}   & 20\%  & 20    & 0.936 & 0.823 & 0.894 & 0.029 & 0.897 & 0.032744314 \\
		\bottomrule
	\end{tabular}%
	\label{tab3}%
\end{table*}%

\section{Discussion}
\label{sec5}
\subsection{Ablation study}
\label{subsec5-1}
To verify the role of the auxiliary models, we combined the main model with two auxiliary models and tested their semi-supervised segmentation performance on the LA dataset. Table \ref{tab4} provides a quantitative comparison of the results of the ablation experiments on the LA dataset. It is easy to see that the semi-supervised segmentation performance of combining the main model with any one of the auxiliary models is significantly improved compared to using the main model alone. When the main model is combined with both auxiliary models, the semi-supervised segmentation performance is the best. This is because the perturbations of the two auxiliary models constructed in this paper complement each other and can help the main model learn more effective information from unlabeled data through consistency training.

\begin{table*}[htbp]
	\centering
	\caption{Ablation studies on the LA dataset. Main denotes the main model, Aux1 denotes the first auxiliary model, and Aux2 denotes the second auxiliary model.}
	\begin{tabular}{llllllll}
		\toprule
		\multicolumn{2}{l}{\#Scans used} & \multicolumn{3}{l}{Models used} & \multicolumn{3}{l}{Metrics} \\
		\midrule
		Labeled & Unlabeled & \multicolumn{1}{l}{Main} & \multicolumn{1}{l}{Aux1} & \multicolumn{1}{l}{Aux2} & Dice(\%)↑ & 95HD(voxel)↓ & ASD(voxel)↓ \\
		\midrule
		8(10\%) & 72(90\%) & \checkmark &       &       & 87.18 & 13.62 & 3.97 \\
		8(10\%) & 72(90\%) & \checkmark & \checkmark &       & 89.75 & 7.39  & 1.95 \\
		8(10\%) & 72(90\%) & \checkmark &       & \checkmark & 88.4  & 8.67  & 2.17 \\
		8(10\%) & 72(90\%) & \checkmark & \checkmark & \checkmark & \textbf{89.82} & \textbf{7.03} & \textbf{1.81} \\
		\midrule
		\midrule
		16(20\%) & 64(80\%) & \checkmark &       &       & 89.38 & 9.64  & 2.81 \\
		16(20\%) & 64(80\%) & \checkmark & \checkmark &       & 91.07 & 6.11  & 1.59 \\
		16(20\%) & 64(80\%) & \checkmark &       & \checkmark & 90.61 & 6.56  & 1.61 \\
		16(20\%) & 64(80\%) & \checkmark & \checkmark & \checkmark & \textbf{91.27} & \textbf{5.75} & \textbf{1.54} \\
		\bottomrule
	\end{tabular}%
	\label{tab4}%
\end{table*}%

\subsection{Effects of independent encoders}
\label{subsec5-2}
Table \ref{tab5} provides a comparison of test results between shared encoder and independent encoder in the LA dataset. It can be seen that the segmentation results are much lower if the auxiliary model encoder is shared, compared to having an independent encoder. Although sharing the encoder can reduce network complexity and training time, it also standardizes the encoder parameters. Even if complementary and symmetric skip connections are used, the pseudo-labels generated by the auxiliary model have almost no useful information for the main model to learn from. This ultimately leads to a significant decrease in the performance of the auxiliary model after sharing the encoder.
\begin{table*}[htbp]
	\centering
	\caption{Performance on the LA dataset comparison between shared and standalone encoders}
	\begin{tabular}{lllllll}
		\toprule
		\multirow{2}[4]{*}{Encoder} & \multicolumn{2}{l}{\#Scans used} & \multicolumn{4}{l}{Metrics} \\
		\cmidrule{2-7}          & Labeled & Unlabeled & Dice(\%)↑ & Jaccard(\%)↑ & 95HD(voxel)↓ & ASD(voxel)↓ \\
		\midrule
		Shared & 8(10\%) & 72(90\%) & 84.3  & 73.67 & 16.99 & 4.79 \\
		Independent  & 8(10\%) & 72(90\%) & \textbf{89.82} & \textbf{81.60} & \textbf{7.03} & \textbf{1.81} \\
		\midrule
		\midrule
		Shared & 16(20\%) & 64(80\%) & 87.9  & 78.65 & 9.99  & 2.45 \\
		Independent & 16(20\%) & 64(80\%) & \textbf{91.27} & \textbf{84.02} & \textbf{5.75} & \textbf{1.54} \\
		\bottomrule
	\end{tabular}%
	\label{tab5}%
\end{table*}%

\subsection{Effects of sharpening temperature $T$}
\label{subsec5-3}
\begin{figure*}[htbp]
	\centering
	\includegraphics[scale=.75]{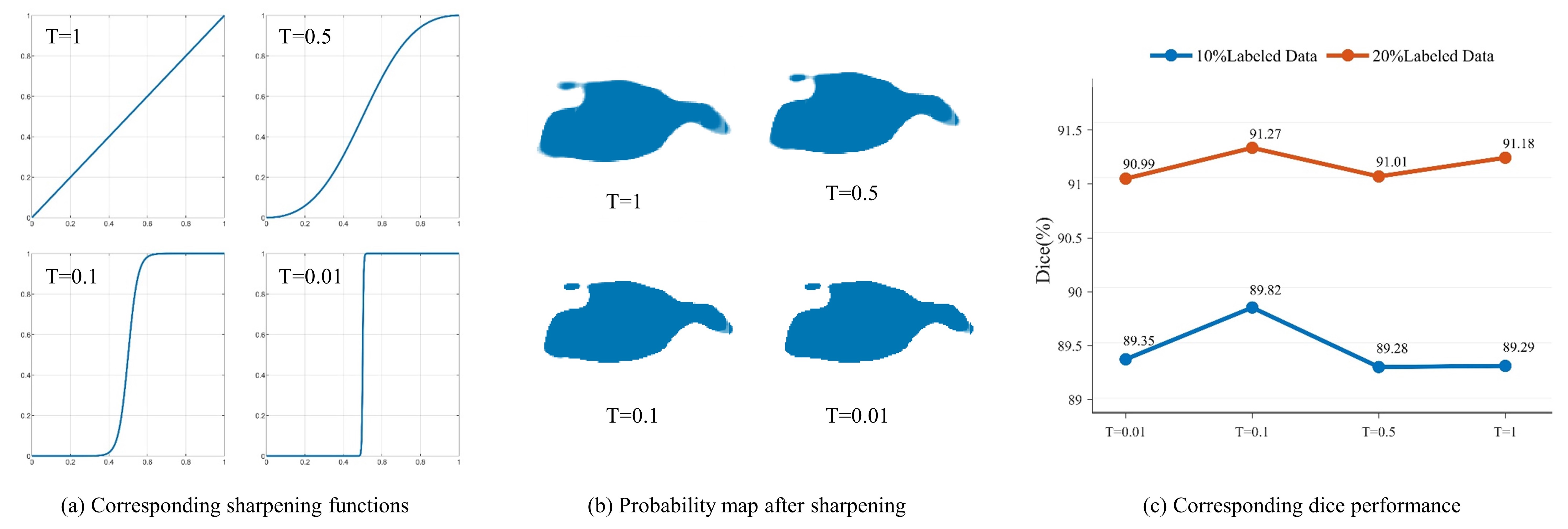}
	\caption{(a)Corresponding sharpening functions, (b)probability map after sharpening, (c)dice performance with different sharpening temperatures T on the LA dataset.}
	\label{fig7}
\end{figure*}

The sharpening temperature $T$ is used to adjust the sharpening function. A smaller $T$ enhances the entropy minimization constraint, but also introduces more noise (as shown in Figure \ref{fig7}(a, b)). To select a suitable $T$, we conducted experiments with different T values on the LA dataset, and the results are shown in Figure \ref{fig7}(c). It can be seen that the temperature function $T=0.1$ can balance the entropy minimization constraint and noise well. Therefore, we finally adopted the sharpening function with $T=0.1$ to generate pseudo-labels for all datasets.

\subsection{Model selection for inference}
\label{subsec5-4}
\begin{figure*}[htbp]
	\centering
	\includegraphics[scale=.7]{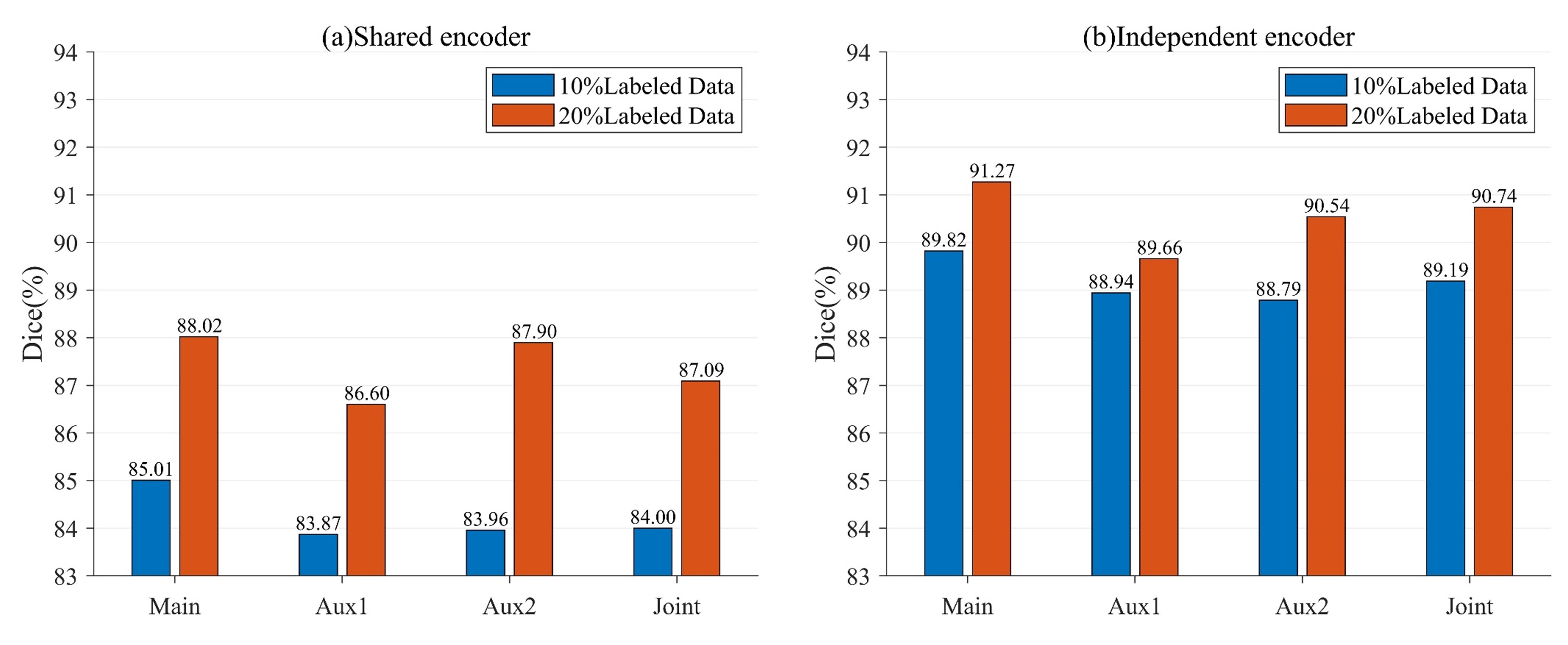}
	\caption{The inference results on the LA dataset under (a) the shared encoder setting and (b) the independent encoder setting for different models. The main model is denoted as Main, Auxiliary model 1 is denoted as Aux1, Auxiliary model 2 is denoted as Aux2, and Joint denotes the joint inference of the main model and the two auxiliary models.}
	\label{fig8}
\end{figure*}

Figure \ref{fig8} provides the inference results of different models on the LA dataset under the settings of shared encoder and independent encoder. It can be observed that the joint inference of the main model and two auxiliary models does not perform as well as the main model alone. It is well known that joint inference takes more inference time than the main model alone. This is because the auxiliary models in this method have some noise, while the main model learns more effective information and is more robust. Therefore, we ultimately choose to use the main model alone for inference.

\subsection{Effects of the weight for the supervised loss ${{\lambda }_{s}}$}
\label{subsec5-5}
\begin{figure}[htbp]
	\centering
	\includegraphics[scale=.7]{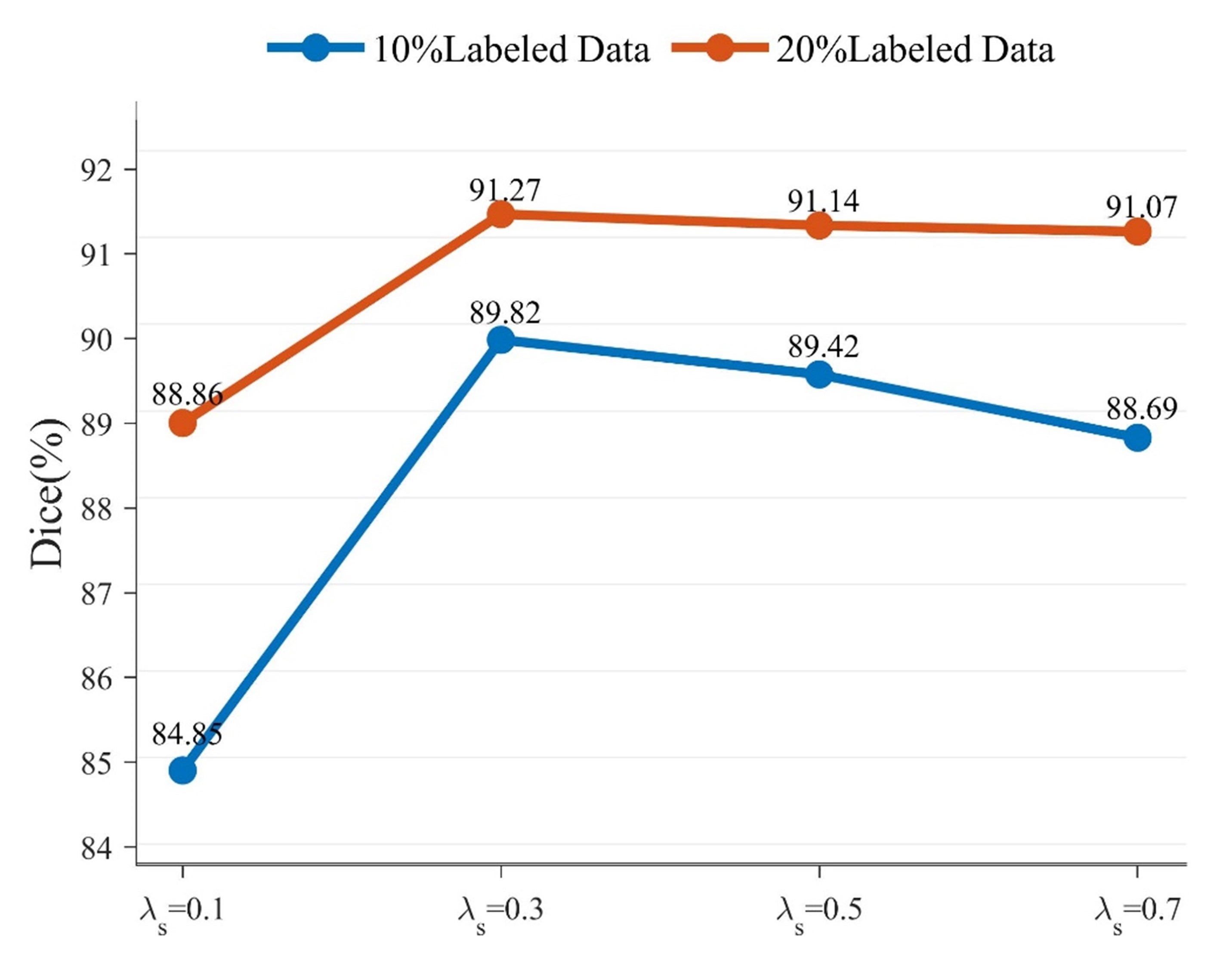}
	\caption{(a)Corresponding sharpening functions, (b)probability map after sharpening, (c)dice performance with different sharpening temperatures T on the LA dataset.}
	\label{fig9}
\end{figure}

This article conducted experiments to select the optimal weight for the supervised loss. Figure \ref{fig9} provides a comparison of the test results with different values of ${{\lambda }_{s}}$. When the value of ${{\lambda }_{s}}$ is too small, the model learns less reliable information from the labeled data, resulting in poorer segmentation performance. If the value of ${{\lambda }_{s}}$ is too large, the consistency training effect will be reduced, leading to a decrease in the segmentation performance of the model. Finally, this article selects ${{\lambda }_{s}}=0.3$ as the weight for the supervised loss.

\subsection{Limitations and future work}
\label{subsec5-6}
Although the method proposed in this paper has excellent semi-supervised segmentation ability, the complementary model perturbation construction method based on the alternating use of skip connections limits the further extension of this method. Additionally, this paper did not consider complementary data-level perturbations. Complementary data-level perturbations may bring better semi-supervised segmentation performance. Future work will focus on the construction method of complementary data-level perturbations and the use of more diverse datasets to validate the proposed method.

\section{Conclusion}
\label{sec6}
This paper proposes a novel semi-supervised learning method based on complementary consistency for 3D left atrial image segmentation. The method constructs complementary model perturbations from a complementary information perspective, and efficiently utilizes unlabeled data through consistency training with complementary information. Two complementary auxiliary models are constructed, one focusing on high-resolution detail information and the other on high-level semantic information. These auxiliary models generate pseudo-labels with complementary information, which are used in consistency training with the main model. During training, the complementary auxiliary models are encouraged to help the main model generate high-quality segmentation results, while the main model is encouraged to help the auxiliary models generate more reliable pseudo-labels. Compared with current advanced methods, CC-Net, proposed in this paper, achieves the best semi-supervised segmentation performance on two datasets.

\section*{Acknowledgments}
The research has been supported by the Natural Science Foundation of Hunan Province - Youth Project (Project No. 2020JJ5201), the Excellent Youth Project of Education Department (Project No. 21B0456) and the Shenzhen Natural Science Foundation Project (Ref. JCYJ20210324101215039).
\bibliographystyle{model2-names.bst}\biboptions{authoryear}
\bibliography{refs}

\end{document}